\documentclass[prb,twocolumn,aps,superscriptaddress,showpacs,floatfix]{revtex4}

\usepackage{amsmath}
\usepackage{amssymb}
\usepackage{amsbsy}
\usepackage{graphicx}
\usepackage{color}
\usepackage{epstopdf}
\usepackage{mathrsfs}

\begin{document}
	
	\title{Topological Bose-Mott insulators in one-dimensional non-Hermitian superlattices}
	
	\author{Zhihao Xu}
    \email{xuzhihao@sxu.edu.cn}
    \affiliation{Institute of Theoretical Physics, Shanxi University, Taiyuan 030006, China}
    \affiliation{Beijing National Laboratory for Condensed Matter Physics,	Institute of Physics, Chinese Academy of Sciences, Beijing 100190, China}
	\affiliation{Collaborative Innovation Center of Extreme Optics, Shanxi University, Taiyuan 030006, P.R.China}
	\affiliation{State Key Laboratory of Quantum Optics and Quantum Optics Devices, Institute of Opto-Electronics, Shanxi University, Taiyuan 030006, P.R.China}	
	
	\author{Shu Chen}
	\email{schen@iphy.ac.cn}
	\affiliation{Beijing National Laboratory for Condensed Matter Physics, Institute of Physics, Chinese Academy of Sciences, Beijing 100190, China}
	\affiliation{School of Physical Sciences, University of Chinese Academy of Sciences, Beijing, 100049, China}
	\affiliation{Yangtze River Delta Physics Research Center, Liyang, Jiangsu 213300, China}

\begin{abstract}
	We study the topological properties of Bose-Mott insulators in one-dimensional non-Hermitian superlattices, which may serve as effective Hamiltonians for cold atomic optical systems with either two-body loss or one-body loss. We find that in the strongly repulsive limit, the Mott insulator states of the Bose-Hubbard model with a finite two-body loss under integer fillings are topological insulators characterized by a finite charge gap, nonzero integer Chern numbers, and nontrivial edge modes in a low-energy excitation spectrum under an open boundary condition. The two-body loss suppressed by the strong repulsion results in a stable topological Bose-Mott insulator which has shares features similar to the Hermitian case. However, for the non-Hermitian model related to the one-body loss, we find the non-Hermitian topological Mott insulators are unstable with a finite imaginary excitation gap. Finally, we also discuss the stability of the Mott phase in the presence of two-body loss by solving the Lindblad master equation.
\end{abstract}
	
\maketitle
\section{Introduction}
Non-Hermitian topological systems have attracted much attention in recent years with the aid of the fast development
of topological photonics \cite{Ozawa,Lu,Guo,Midya,Xu,Ding,Ghatak,Alvarez1}. Since non-Hermitian systems possess more fundamental nonspatial symmetries, their topological classification goes beyond the standard ten classes of the corresponding Hermitian systems \cite{Gong,LiuCH1,LiuCH2,Sato,Zhou,Lieu-BDG,Karabata,SongZhi}. Non-Hermitian systems have been shown to exhibit many exotic properties without Hermitian counterparts \cite{Hatano,Levitov,Esaki,ZhuBG,ZhihaoXu,Yuce2015,XuY,Leykam,ShenH,TELee,Yin,Lieu,JiangHui,Yao,Alvarez,Xiong,Kunst,TaoLiu2019,ZiYongGe,NoriNC}, including half-integer topological invariants \cite{TELee,Yin,Lieu,JiangHui,ShenH,Leykam},
the non-Hermitian skin effect, and breakdown of bulk-boundary correspondence in some nonreciprocal systems \cite{Yao,Alvarez,Xiong,Gong,Kunst,Yao1,Jin,Lee,Kawabata,Lee1,WangZhong2019,Herviou,Kou,Rui,Xue,Longhi,JiangHui2019,Turker,Yokomizo,Fang,HuJP,Okuma,Borgnia}. Recently, exploring topological phases in interacting non-Hermitian systems was also addressed in several works \cite{ChenWQ,NH-FQH,KYamamoto}.

Due to their highly controllability, one-dimensional (1D) optical superlattices, which can be realized by superimposing two 1D optical lattices with commensurate wavelengths, have provided an ideal playground for exploring topologically nontrivial phases \cite{LangLJ,Kraus,Ganeshan,GuoHM,XuZH,XuZH2,Kuno,HuHP,ZhuSL,Yoshida,ShengDN,Grusdt}. The interplay between many-body interaction and single-particle band topology can lead to intriguing correlated states exhibiting nontrivial topological properties, e.g., fractional topological states \cite{XuZH,ShengDN} and topological Mott insulators (TMIs) \cite{XuZH2,Kuno,HuHP,ZhuSL,Yoshida,Grusdt} in interacting superlattice systems. Recent experimental progress demonstrates that optical lattice systems may be a controllable candidate for studying quantum open systems by introducing a dissipation process \cite{LuoL,Patil2015,Bouganne,SchneiderPRX} and the exciton-polaritons releasing coherent radiation are an open quantum system requiring constant pumping of energy and continuously decaying \cite{TGao2015,Hanai,TGao2018}, which can be effectively described by an effective non-Hermitian Hamiltonian under certain conditions.
So far, different kinds of studies on the effect of the dissipation have been reported due to the particle loss or photon scattering \cite{DWitthaut,FVerstraete,YJHan,EGDallaTorre,SDiehl1,SDiehl2,ATomadin,CCiuti,WHofstetter,KStannigel,SFurukawa,NDogra}. Especially, the two-body loss on quantum many-body systems is realized in the form of inelastic collisions which can be widely controlled \cite{NSyassen,AHazzard,BGadway,TTomita,Sponselee}. One can tune the inelastic two-body scattering via a photo-association resonance in a bosonic or fermionic system, where a delay of the melting of the Mott insulating state was detected \cite{TTomita,Sponselee}.

Motivated by this progress, in this work we study the realization of the topological Mott phase in 1D non-Hermitian superlattice systems. For an interacting bosonic gas trapped in a superlattice with an integer band filling factor, a TMI is in the formation of the Mott phase characterized by a nontrivial topological invariant \cite{ZhuSL,XuZH2}. When we consider a non-Hermitian system, its eigenenergies are generally complex, and natural questions that arise here are whether the Mott phase still exists and how to characterize the non-Hermitian TMI.

\section{Model and method}
To address these problems, we first consider interacting bosons trapped in a 1D two-periodic optical lattice with the Hamiltonian described by $H_{\mathrm{BH}}=H_0+H_I$, with
\begin{equation}\label{eq1}
H_0=-\sum_j J(j,\theta)(\hat{b}_j^{\dagger}\hat{b}_{j+1}+\mathrm{H.c.})+\sum_j \frac{V(j,\theta)}{2}\hat{n}_j,
\end{equation}
and
\begin{equation}\label{eq2}	
H_I=\frac{U}{2}\sum_j \hat{n}_j(\hat{n}_j-1).
\end{equation}
Here, $\hat{b}_j$ is the annihilation operator of the bosons at site $j$, $\hat{n}_j=\hat{b}^{\dagger}_j\hat{b}_j$ is the number operator of bosons, and the alternating hopping strengths are given by $J(j,\theta)=J[1+ \delta \sin{(\pi j+\theta)}]$ with the dimerization strength $\delta$ and $J$ being set as the energy unit ($J=1$). The on-site potential is given by $V(j,\theta)=V\cos{(\pi j+\theta)}$, with $V\cos{\theta}$ denoting the energy offset between neighboring sites for the real phase $\theta$ of the potential. We carefully choose the strength of the alternating tunneling term $J(j,\theta)$ and the strength of the chemical potentials $V(j,\theta)$ to keep the gap of the two-band model open during the rolling of $\theta$, which is necessary to form a topological insulator. The interaction strength $U=U_r$ is always real and can be experimentally controlled by the Feshbach resonance. Complex-valued interactions can emerge in some effective Hamiltonians of ultracold atomic systems induced by considering the inelastic processes between different orbitals which give rise to two-particle loss.

When atoms undergo inelastic collisions, the scattered atoms are lost from the system. The atom losses are described by a quantum master equation:
\begin{align}\label{eq3}
\partial_t\rho(t)=&-i[H,\rho(t)]-\frac{1}{2}\gamma\sum_{j}[L_j^{\dagger}L_j\rho(t)+\rho(t)L_j^{\dagger}L_j \notag \\
&-2L_j\rho(t)L_j^{\dagger}],
\end{align}
where $\rho(t)$ is the density matrix of the atomic gas and $L_j$ is a Lindblad operator at site $j$ which describes a loss with the rate $\gamma>0$.  The process of two-particle loss can be described by setting $L_j\to \hat{b}_j\hat{b}_j$. Considering the short time evolution, the quantum-jump term can be negligible, and the dynamical evolution is described by
\begin{equation}\label{eq4}	\partial_t\rho(t)=-i[H^{(1)}_{\mathrm{eff}}\rho(t)-\rho(t)H^{(1)}_{\mathrm{eff}}],
\end{equation}
where $H^{(1)}_{\mathbf{eff}}=H_0+H_I$ is an effective Hamiltonian, with the interaction amplitude $U$ taking a complex value,
\begin{equation}\label{eq5}
U = U_r-i\gamma.
\end{equation}
Here, $U_r \ge 0$ represents the repulsive interaction, and $\gamma\ge 0$ is the rate of loss. In such a case, the lowest real part of the spectrum is the effective ground state, and the imaginary part of the energy denotes the decay rate of each eigenstate.

\begin{figure}[tbp]
	\begin{center}
		\includegraphics[width=.5 \textwidth] {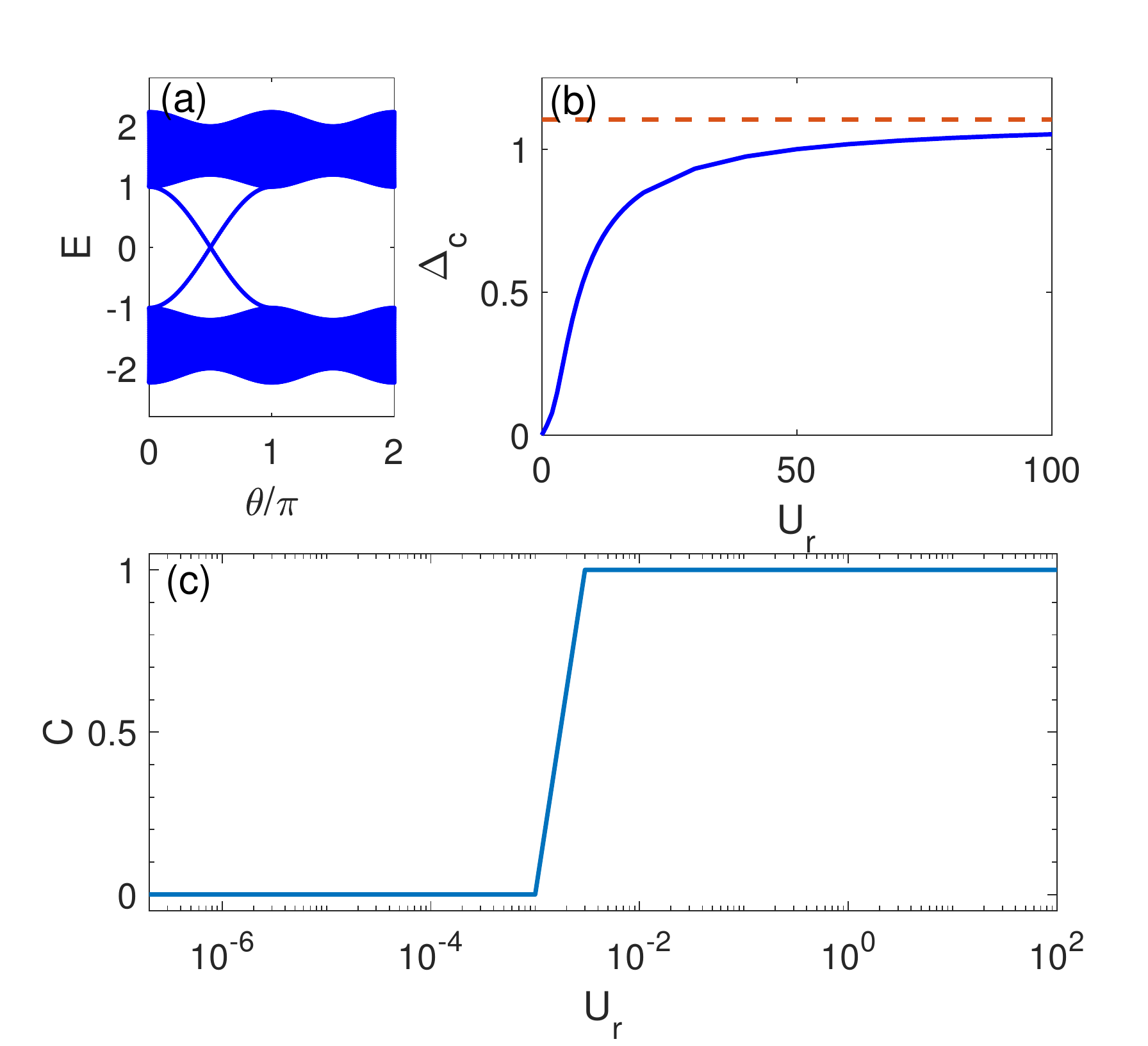}
	\end{center}
	\caption{(Color online)  (a) Single-particle energy spectrum of the Rice-Mele model as a function of $\theta$ under the OBC. (b) The charge gap $\Delta_c$ as a function of $U_r$ for the Hermitian case with $\theta=\pi/4$, $L=14$, and $\nu=1$ under the PBC. The dashed line denotes $\Delta_b/2=1.1045$. (c) The Chern numbers versus $U_r$ with $\nu=1$. Here, $\delta=0.6$ and $V=2$. }\label{Fig1}
\end{figure}

When $U=0$, the Hamiltonian reduces to the topologically nontrivial  Rice-Mele model \cite{Rice-Mele}. The energy spectrum in momentum space with momentum $k$ is $\varepsilon_{\pm}=\pm\sqrt{2+V^2\cos^2\theta/4+2\delta^2\sin^2\theta+2(1-\delta^2\sin^2\theta)\cos{k}}$, and the energy gap is $\Delta_b=2\sqrt{V^2\cos^2\theta/4+4\delta^2\sin^2\theta}$. 
Figure \ref{Fig1}(a) shows the single-particle energy spectrum of the Rice-Mele model as a function of $\theta$ with $\delta=0.6$, $V=2$, and $\theta=\pi/4$ under the open boundary condition (OBC). As the phase shift $\theta$ varies from zero to $2\pi$, the edge states connecting two bulk spectra emerge in the gap regime which characterizes the topologically nontrivial feature. For a Fermi system, when the band filling factor $\nu=N/N_{cell}$ is an integer, with $N$, $L$, and $N_{cell}=L/2$ representing the number of particles, the lattice site, and the unit cell of  the system, respectively, it corresponds to a band insulator with the lowest $\nu$ bands being fully filled by the fermions. Such a Fermionic system was demonstrated to be topologically nontrivial with a nontrivial topological number. However, for a noninteracting bosonic case, all the bosons are condensed in the $k=0$ state, and the system is in a superfluid state with a trivial topological number.

When the repulsive interaction $U_r$ is considered, the bosonic system with an integer band filling factor $\nu$ can enter a Mott phase. We can calculate the charge gap defined as
\begin{equation}\label{eq6}
\Delta_c=\frac{1}{2}[E_0(N+1)+E_0(N-1)]-E_0(N)
\end{equation}
by numerically diagonalizing the Hamiltonian $H_{\mathrm{BH}}$, where $E_0(N)$ represents the ground-state energy of the $N$-boson system. Figure \ref{Fig1}(b) shows the charge gap $\Delta_c$ versus $U_r$ for the Hermitian case $H_{\mathrm{BH}}$ with $L=14$, $\delta=0.6$, $V=2$, $\theta=\pi/4$, and $\nu=1$ under the periodic boundary condition (PBC). In the small-$U_r$ limit, the charge gap $\Delta_c \to 0$, and the charge gap $\Delta_c$ grows with the increase of $U_r$. In the strongly repulsive limit, $\Delta_c$ tends to $\Delta_b/2$, as predicted by applying the Bose-Fermi mapping \cite{XuZH2}. Our results indicate that a Mott insulator phase emerges in the large-$U_r$ case.

The topological feature of the Mott insulator state can be characterized by the many-body Chern number in the two-dimensional (2D) parameter space $(\varphi,\theta)$ \cite{XuZH,XuZH2,ZhuSL}. Here, we introduce the twist boundary condition, which corresponds to a shift momentum $k=(2\pi m+\varphi)/N_{cell}$ in the Brillouin zone, with $\varphi$ being a generalized boundary phase and $m=0,1,\dots,N_{cell}-1$. The many-body Chern number for the Hermitian case is defined as $C=\frac{1}{2\pi}\int d\varphi d\theta B(\varphi,\theta)$, with the Berry curvature $B(\varphi,\theta)=\mathrm{Im}\left(\langle\frac{\partial\psi}{\partial \theta}|\frac{\partial \psi}{\partial \varphi}\rangle - \langle\frac{\partial \psi}{\partial \varphi}|\frac{\partial \psi}{\partial \theta}\rangle \right)$. We apply the exact diagonalization of a finite-size system $L=12$ to obtain the exact value of $C$ with the manifold of the torus discretized by $6\times 6$ meshes.  When the Mott insulator is formed for $\nu=1$, the corresponding state has a Chern number $C=1$, indicating that the Mott insulator is topologically nontrivial, i.e., the formation of a TMI. As shown in Fig. \ref{Fig1}(c), when the system is in the gapless superfluid phase, the Chern number $C$ is unquantized. Our numerical calculations demonstrate that the TMI emerges with the increase of $U_r$, which is consistent with previous studies \cite{XuZH2,ZhuSL}.

\begin{figure}[tbp]
	\begin{center}
		\includegraphics[width=.5 \textwidth] {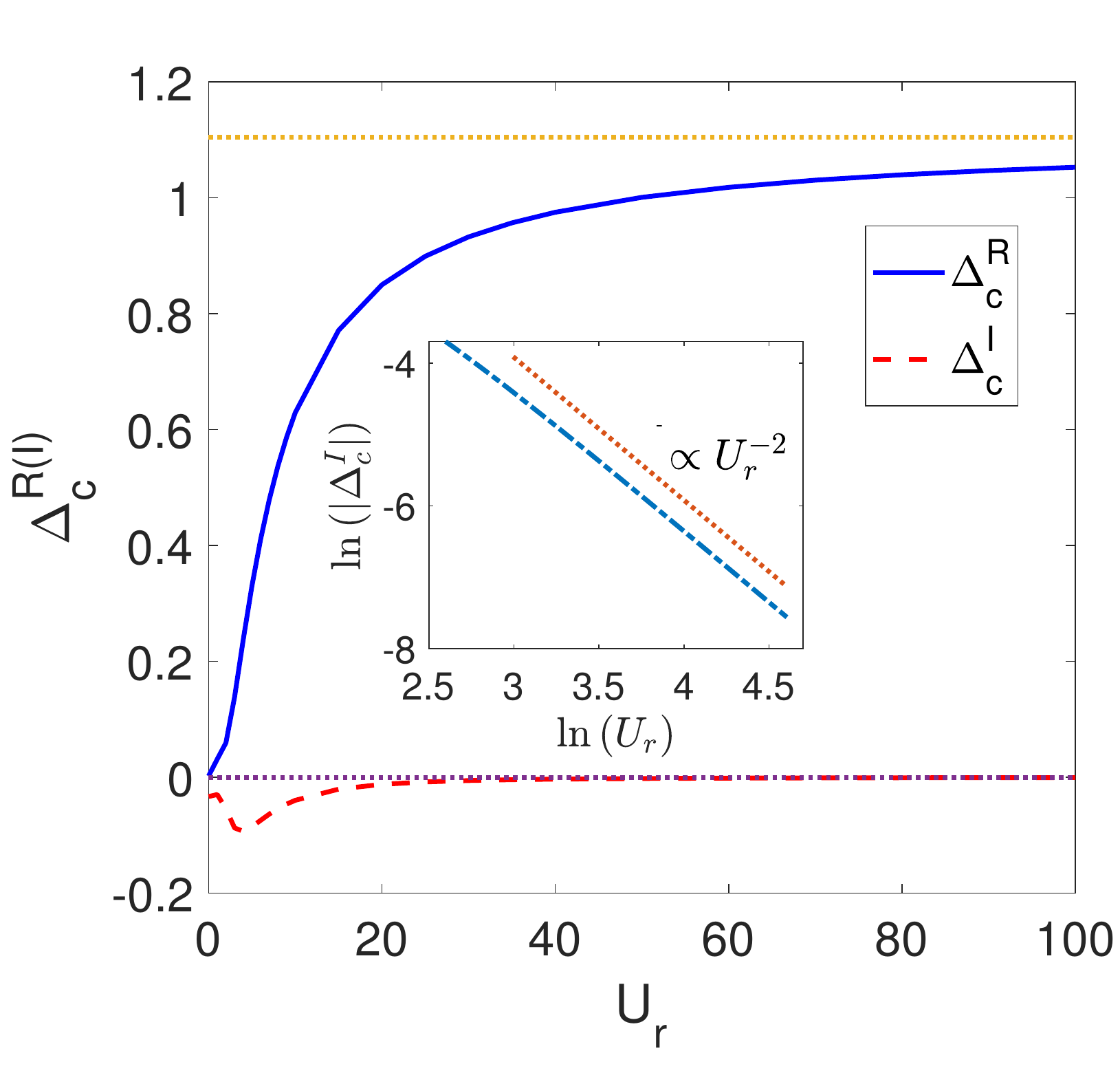}
	\end{center}
	\caption{(Color online) The real part $\Delta_c^{R}$ and the imaginary part $\Delta_c^{I}$ of the charge gap as a function of $U_r$ for the Hamiltonian $H_{\mathrm{eff}}^{(1)}$ with $L=14$, $\nu=1$, $\delta=0.6$, $V=2$, $\theta=\pi/4$, and $\gamma=1$ under the PBC. The orange dotted line denotes $\Delta_b/2=1.1045$, and the value of the purple dotted line is zero. The inset shows the power-law decay of $|\Delta_c^{I}|$ with $U_r$ in the strongly interacting limit, and the dashed line in the inset indicates a power-law fitting.}\label{Fig2}
\end{figure}

\section{Non-Hermitian TMI} 
Now we study the non-Hermitian effect induced by the imaginary part of $U$.
As a concrete example, we consider the system described by Hamiltonian $H^{(1)}_{\mathrm{eff}}$ with $\nu=1$, $\delta=0.6$, $V=2$, $\theta=\pi/4$, and $\gamma=1$ and calculate the charge gap $\Delta_c$ defined by Eq. (\ref{eq6}). Here, due to the complex-energy spectrum for a non-Hermitian system, we define $E_0(N)$ as the ground-state energy for the $N$-boson system with the minimum value of the real part. Correspondingly, the real part of the charge gap for the non-Hermitian case is defined as $\Delta_c^{R}=\mathrm{Re}(\Delta_c)$, and the imaginary part of the charge gap $\Delta_c^{I}=\mathrm{Im}(\Delta_c)$ is also calculated. Figure \ref{Fig2} shows $\Delta_c^{R}$ and $\Delta_c^{I}$ as a function of $U_r$ under the PBC. For $U_r=0$, the real part of the charge gap $\Delta_c^{R} \to 0$, and $\Delta_c^{I}$ takes a finite value.
The real part of the charge gap $\Delta_c^{R}$ shows a monotonic increase with the increase of $U_r$ and tends to $\Delta_b/2$ in the strongly repulsive limit. However, with the increase of $U_r$, the absolute value of $\Delta_c^{I}$ first increases and then decreases when $U_r>3.76$. In the large-$U_r$ limit, according to the second perturbation theory, $\Delta_c^{I} \propto -\gamma/U_r^2$ presents a power-law decay with the increase of $U_r$, which is shown in the inset of Fig. \ref{Fig2}. The two-body collisions are strongly suppressed, which leads to the decrease of the atomic loss, indicating that a stable Mott insulator emerges in the strongly interacting limit.

To characterize the topological feature, we now construct the many-body Chern number for the non-Hermitian case. In the 2D parameter space of $(\varphi,\theta)$, the Chern number of the ground state for a non-Hermitian system is an integral invariant, which can be defined as
\begin{equation}\label{eq7}	
C^{\alpha\beta}=\frac{1}{2\pi}\int d\varphi d\theta B^{\alpha\beta}(\varphi,\theta),
\end{equation}
where the Berry curvatures
$B^{\alpha\beta}(\varphi,\theta)=\mathrm{Im}\left(\langle\frac{\partial\psi^{\alpha}}{\partial \theta}|\frac{\partial \psi^{\beta}}{\partial \varphi}\rangle - \langle\frac{\partial \psi^{\alpha}}{\partial \varphi}|\frac{\partial \psi^{\beta}}{\partial \theta}\rangle \right)$, with $\alpha,\beta=R/L$. These definitions are a direct generalization of non-Hermitian Chern numbers \cite{ShenH} to the many-body systems. We note that for an open quantum system the topological invariant can be extracted from the ensemble geometric phase for mixed quantum states \cite{BardynPRX}. Here, the right eigenstates $|\psi^{R}\rangle$ and the left eigenstates $|\psi^{L}\rangle$ can be respectively defined as $H|\psi^{R}\rangle=E|\psi^{R}\rangle$ and $H^{\dagger}|\psi^{L}\rangle=E^{*}|\psi^{L}\rangle$, with the normalization condition $\langle \psi^{\alpha}|\psi^{\beta}\rangle=1$. We numerically calculate four different Chern numbers, $C^{LL}$, $C^{LR}$, $C^{RL}$, and $C^{RR}$, of the Hamiltonian $H_{\mathrm{eff}}^{(1)}$ with $\nu=1$, $\gamma=1$, and different $U_r$. We find $C^{LL}=C^{LR}=C^{RL}=C^{RR}=1$ for this non-Hermitian interacting boson system with finite repulsive $U_r$. The proof of the equivalence of four many-body Chern numbers for non-Hermitian cases is presented in Appendix A. In the large-$U_r$ limit, the stable Mott insulators are formed for $\nu=1$, corresponding to the states with nontrivial Chern numbers, which show features similar to the Hermitian case, and the stable Mott insulators are topologically nontrivial.
\begin{figure}[tbp]
	\begin{center}
		\includegraphics[width=.5 \textwidth] {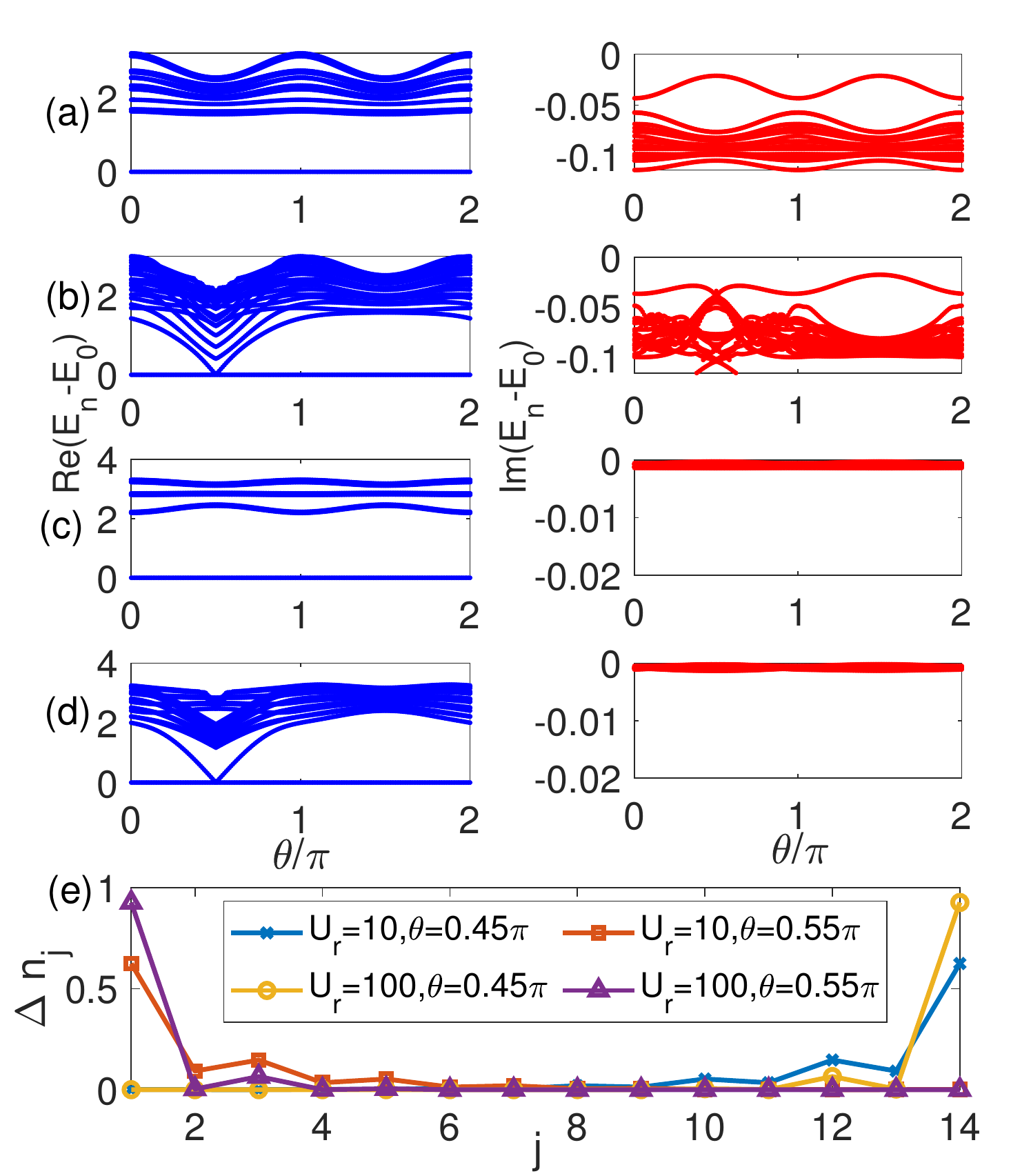}
	\end{center}
	\caption{(Color online) (a)-(d) The real part (left column) and the imaginary part (right column) of the low-energy spectrum $E_n-E_0$ versus $\theta$ with different $U_r$, $L=12$, $\nu=1$, $\delta=0.6$, $V=2$, and $\gamma=1$. (a) $U_r=10$ under the PBC. (b) $U_r=10$ under the OBC. (c) $U_r=100$ under the PBC. (d) $U_r=100$ under the OBC. (e) The density distributions of the edge modes $\Delta n_j$ with $L=14$, $N=7$, $\delta=0.6$, $V=2$, $\gamma=1$, and different $U_r$ and $\theta$ under the OBC.}\label{Fig3}
\end{figure}

According to the bulk-edge correspondence, one may expect that the non-Hermitian TMIs also exhibit nontrivial edge states under the OBC. In Fig. \ref{Fig3}, we show the real part (left column) and the imaginary part (right column) of the low-energy spectrum $E_n-E_0$ versus $\theta$ with $\nu=1$, $\gamma=1$, $L=12$, and various $U_r$ under both the PBC and the OBC, where $n$ marks the energy states to fulfill $\mathrm{Re}(E_1)\le \mathrm{Re}(E_2)\le \cdots$. As shown in Figs. \ref{Fig3}(a) and \ref{Fig3}(c), there is an obvious gap of the real part of the low-energy spectrum between the ground state and the first excited state for systems with $U_r=10$ and $U_r=100$ under the PBC, respectively. Correspondingly, the edge states emerge in the real gap regimes of the low-energy excitation spectrum under the OBC. As the phase shift $\theta$ varies from zero to $2\pi$, the edge states connect the ground state to the excited band, as shown in Figs. \ref{Fig3}(b) and \ref{Fig3}(d) for systems with $U_r=10$ and $U_r=100$ under the OBC, respectively. The real part of the low-energy excitation spectrum for the system with $U_r=100$ of $H_{\mathrm{eff}}^{(1)}$ exhibits almost the same behaviors as its corresponding Hermitian case. The imaginary part of low-energy excitation spectrum gradually converges to zero with the growth of $U_r$, indicating that the effect of the finite two-body loss is almost completely suppressed and a stable TMI exists in the strongly repulsive limit. To analyze the edge modes in the strongly interacting limit, we numerically calculate the density distributions of such in-gap modes which can be defined as $\Delta n_j=\langle \psi^{R}_{N+1}|\hat{n}_j |\psi^{R}_{N+1}\rangle-\langle \psi^{R}_{N}|\hat{n}_j |\psi^{R}_{N}\rangle$, where $|\psi^{R}_{N}\rangle$ is the ground-state wave function of the right state with $N$ bosonic atoms. Figure \ref{Fig3}(e) shows the density distributions of the edge modes with $L=14$, $N=7$, $\delta=0.6$, $V=2$, and $\gamma=1$ under the OBC. The in-gap states mainly distribute near the right edge for $\theta=0.45\pi$ and both $U_r=10$ and $100$. With the increase of $\theta$, there are bosons shifting from the right edge to the left in the large-$U_r$ case as shown in Fig. \ref{Fig3}(e) with $\theta=0.55\pi$. With the rolling of $\theta$ in the large-repulsion case, the boson pumping from one edge to another indicates the topologically nontrivial property of the Bose-Mott insulator phase.

\begin{figure}[tbp]
	\begin{center}
		\includegraphics[width=.5 \textwidth] {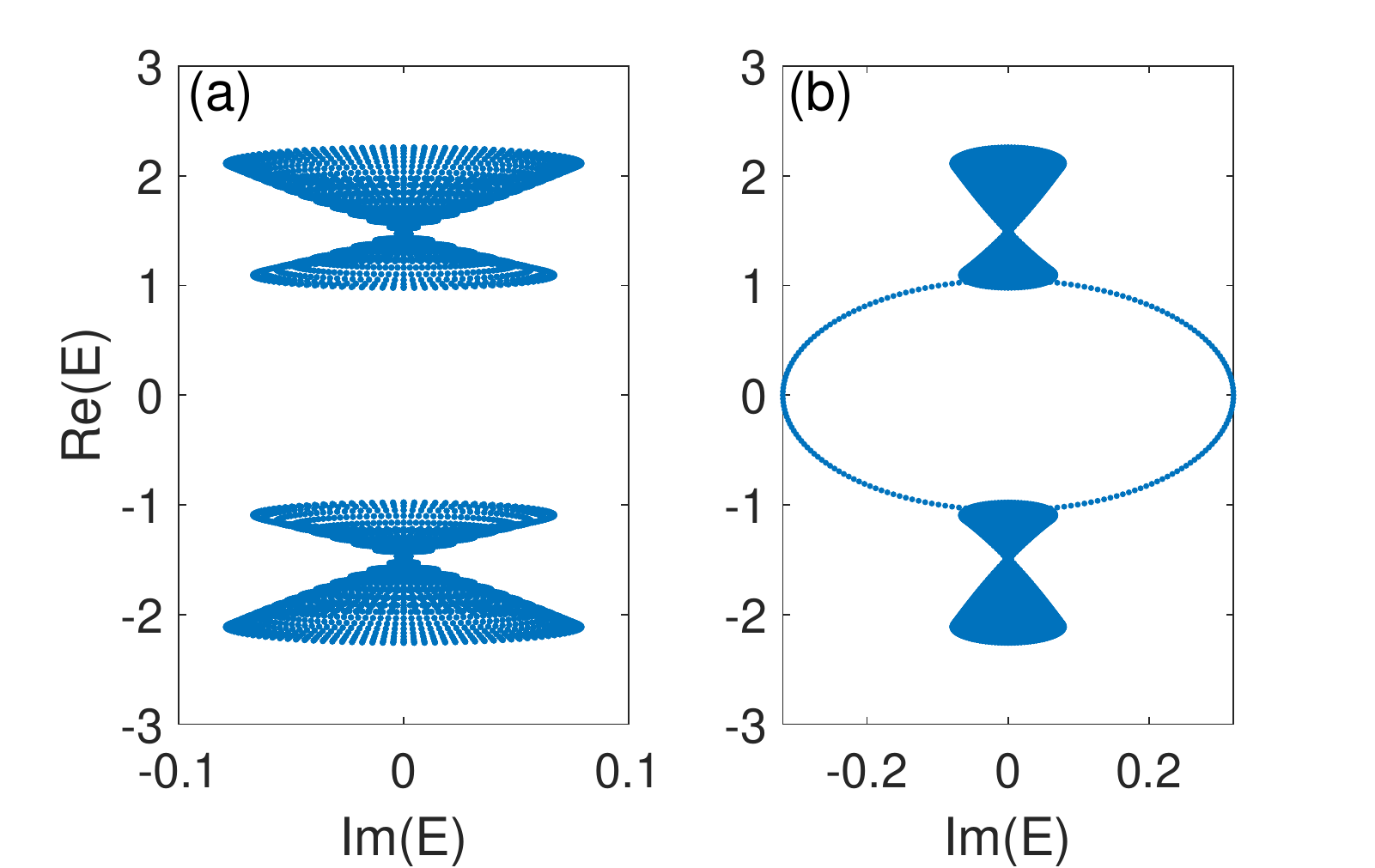}
	\end{center}
	\caption{(Color online) Complex-energy spectrum of $H^{\mathrm{NH}}_{0}$ by rolling $\theta\in[0,2\pi]$ with $\beta=0.1\pi$, $\delta=0.6$, and $V=2$ under (a) the PBC and (b) the OBC.}\label{Fig4}
\end{figure}


\section{TMI in the non-Hermitian Rice-Mele model}
Next, we consider another non-Hermitian extension of the Hamiltonian by taking a complex phase $\theta$, i.e.,  $H^{(2)}_{\mathbf{eff}}=H_0^{\mathrm{NH}}+H_I$, where  $H_0^{\mathrm{NH}}$ is obtained by replacing $\theta \to \theta+i\beta$ in $H_0$, with $\beta$ being an imaginary phase shift. The non-Hermitian Hamiltonian $H_0^{\mathrm{NH}}$ may serve as an effective Hamiltonian related to some one-body loss processes \cite{WangZong-2019,Wunner}, and similar models were studied in Refs. \cite{JiangHui-PRB,Longhi-PRL,ZengQB}. We take $\beta=0.1\pi$ as an example to exhibit our calculation results. In the absence of $U_r$, the energy gap in the complex-energy plane as defined in Ref. \cite{ShenH} emerges for rolling $\theta\in[0,2\pi]$, which is shown in Fig. \ref{Fig4}(a) with $\delta=0.6$ and $V=2$ under the PBC. When the OBC is considered, the edge states emerge between the two bulk bands [see Fig. \ref{Fig4}(b)], with the corresponding bulk state characterized by a nonzero band Chern number \cite{ShenH} .

\begin{figure}[tbp]
	\begin{center}
		\includegraphics[width=.5 \textwidth] {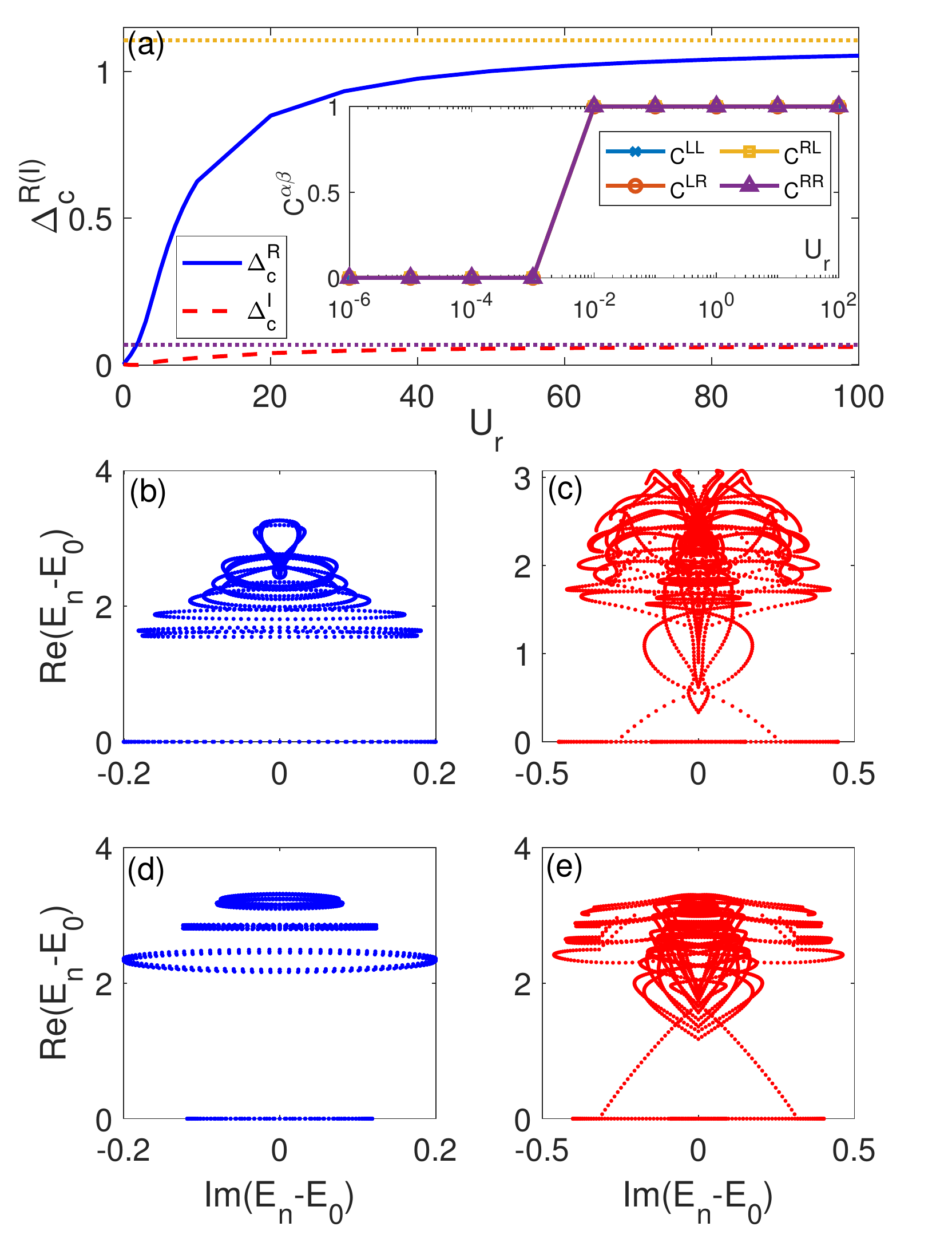}
	\end{center}
	\caption{(Color online) (a) The real part $\Delta_c^{R}$ and the imaginary part $\Delta_c^{I}$ of the charge gap as a function of $U_r$ for the Hamiltonian $H_{\mathrm{eff}}^{(2)}$ with $\theta=\pi/4$ under the PBC. The value of the orange dotted line is equal to $\mathrm{Re}(\Delta_b)/2=1.1065$, and the value of the purple dotted line amounts to $\mathrm{Im}(\Delta_b)/2=0.0667$. (b)-(e) Low-energy excitation spectra of $H_{\mathrm{eff}}^{(2)}$ by rolling $\theta \in [0,2\pi]$. (b) $U_r=10$ under the PBC. (c) $U_r=10$ under the OBC. (d) $U_r=100$ under the PBC. (e) $U_r=100$ under the OBC. The inset in (a) shows the four Chern numbers $C^{\alpha\beta}$ versus $U_r$. Here, $L=14$, $\nu=1$, $\delta=0.6$, $V=2$, and $\beta=0.1\pi$.}\label{Fig5}
\end{figure}

In the presence of the interaction, we first calculate the charge gap $\Delta_c$ versus $U_r$ with $\beta=0.1\pi$, $\delta=0.6$, $V=2$, $\theta=\pi/4$, and $L=14$ under the PBC, as shown in Fig. \ref{Fig5}(a). With the increase of repulsion, both the real and imaginary parts of the charge gap $\Delta_c^{R}$ increase, and in the strong repulsion, $\Delta_c^{R}\to \mathrm{Re}(\Delta_b)/2=1.1065$, and $\Delta_c^{I} \to \mathrm{Im}(\Delta_b)/2=0.0667$. A finite imaginary-valued gap indicates that the Mott phase will collapse in the long-time evolution. To characterize the topological property of the unstable Mott insulators, we numerically calculate the Chern number equation (\ref{eq7}) defined in the non-Hermitian case. For a finite $U_r$, the four Chern numbers are equal, and  $C^{LL}=C^{LR}=C^{RL}=C^{RR}=1$, as shown in the inset of Fig. \ref{Fig5}(a). Due to the finite-size effect, the Chern number is not well defined for small $U_r$. The existence of the nonzero Chern number implies the repulsively interacting system with an integer filling factor with nontrivial topological properties. The topological phase would exhibit nontrivial edge modes in the gap regions under the OBC. We calculate the low-energy excitation spectra of $H_{\mathrm{eff}}^{(2)}$ with $L=14$, $\nu=1$, $\delta=0.6$, $V=2$, and $\beta=0.1\pi$ by rolling $\theta \in [0,2\pi]$, as shown in Figs. \ref{Fig5}(b)-\ref{Fig5}(e).
As shown in Figs. \ref{Fig5}(b) and \ref{Fig5}(d), an obvious energy gap is seen between the ground state and the low-energy excitation states for $U_r=10$ and $100$ under the PBC, respectively. Under the OBC, as the phase shift $\theta$ varies from zero to $2\pi$, the low-energy spectra with $U_r=10$ and $100$ are shown in Figs. \ref{Fig5}(c) and \ref{Fig5}(e), respectively. The edge states connecting the ground states and low-energy part emerge in the gap, and the position of the edge states continuously varies with the rolling of $\theta$ in the complex-energy plane. Specifically, the energies of the ground states and the edge modes exhibit a finite imaginary part, which implies that the non-Hermitian TMIs  with one-body loss are formed, but due to the finite imaginary parts of the ground energies, the TMIs are unstable and will break down during the time evolution.

\section{Summary and discussion} In summary, we have discussed topological Bose-Mott insulators in 1D non-Hermitian superlattices which are characterized by a finite charge gap, nonzero integer Chern numbers defined in the non-Hermitian case, and nontrivial edge modes in the low-energy excitation spectrum under the OBC. We found that for the non-Hermitian effect induced by a finite two-body loss, the nontrivial TMIs are stable in the strong-repulsion limit. However, for a non-Hermitian TMI associated with one-body loss, the low-energy excitation spectrum with a finite
imaginary part suggests the TMI is unstable and will collapse with time.

As our results are obtained on the framework of the effective Hamiltonian, it is important to ask whether the conclusion of the existence of stable TMIs in the  strong-repulsion limit still holds true if we consider the quantum-jump terms by solving the Lindblad master equation. Although a full understanding of correlated topological states in the scheme of open systems is still a challenging open question \cite{Yoshida2020}, in order to gain insight into the above question,  here, we shall study the dynamical evolution of the full master equation described by Eq. (\ref{eq3}) with the Bose-Hubbard model $H_{\mathrm{BH}}$ subjected to a two-particle loss by applying a quantum-trajectory method \cite{Dalibard,Daley,Plenio,Nakagawa2020,Ashida2020,Dum} (see Appendix B for the description of the method in detail).

\begin{figure}[tbp]
	\begin{center}
		\includegraphics[width=.5 \textwidth] {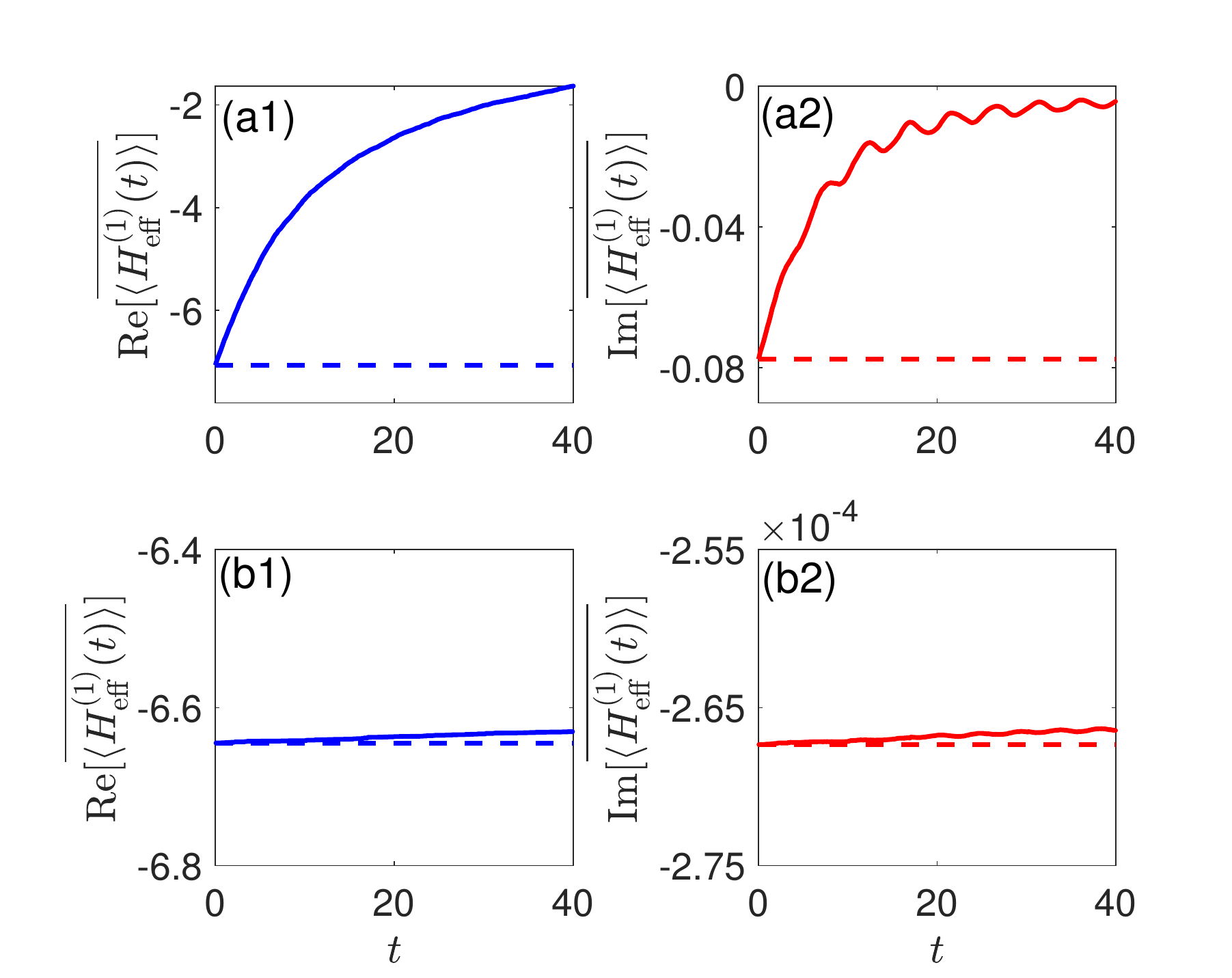}
	\end{center}
	\caption{(Color online) (a1) and (a2) The real and imaginary parts of $\overline{\langle H_{\mathrm{eff}}^{(1)}(t)\rangle}$ for $U_r=5$, respectively. (b1) and (b2) The real and imaginary parts of $\overline{\langle H_{\mathrm{eff}}^{(1)}(t)\rangle}$ for $U_r=100$, respectively.  Here, $J=1$, $L=8$, $\nu=1$, $\theta=\pi/4$, $\delta=0.6$, $V=2$, and $\gamma=1$. The dashed lines correspond to the cases without quantum-jump events.}\label{Fig6}
\end{figure}

\begin{figure}[tbp]
	\begin{center}
		\includegraphics[width=.5 \textwidth] {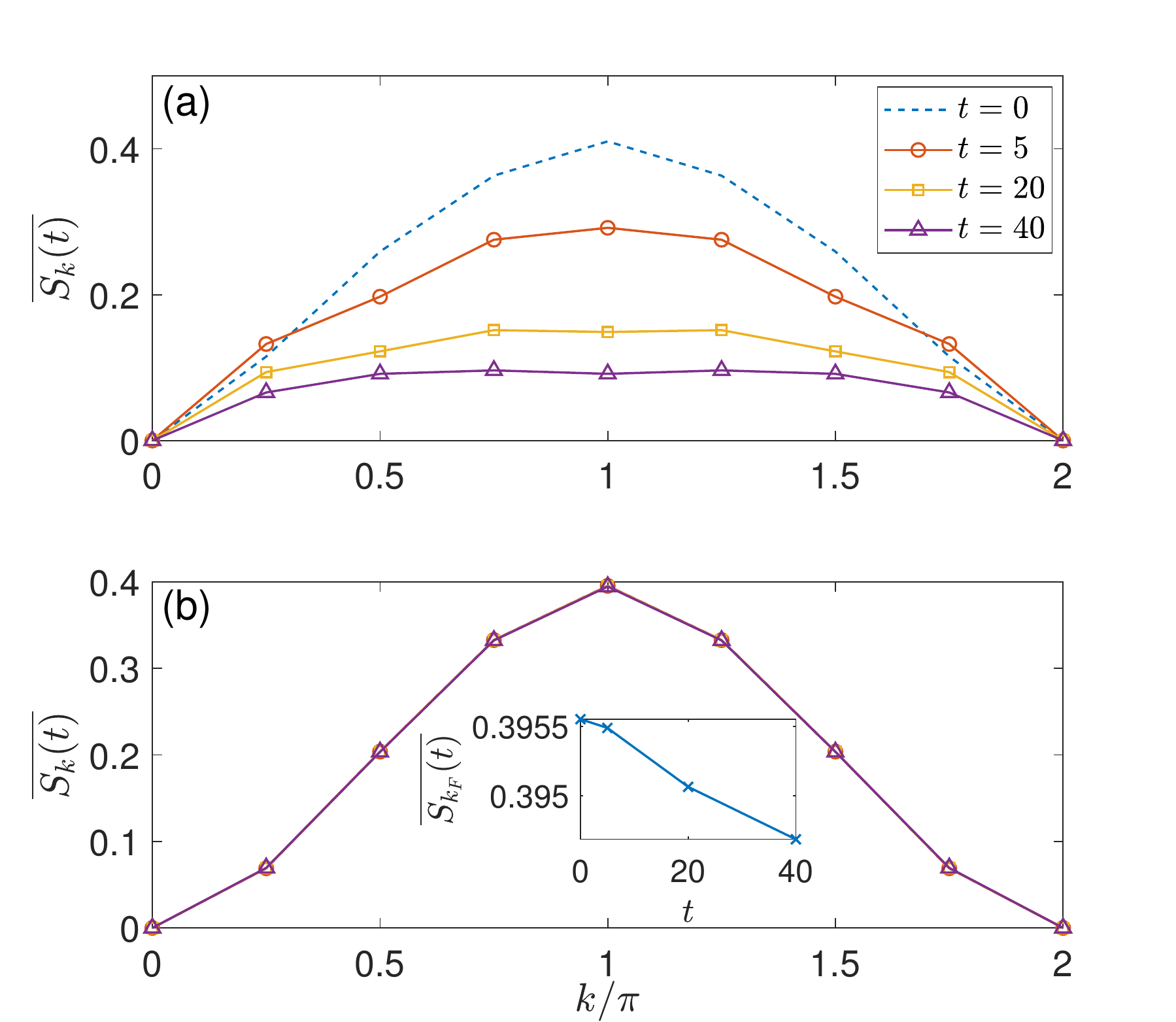}
	\end{center}
	\caption{(Color online) The structure factor $\overline{S_k(t)}$ at different $t$ for (a) $U_r=5$ and (b) $U_r=100$. The inset of (b) shows $\overline{S_{k_F}(t)}$ at different $t$. Here, $J=1$, $L=8$, $\nu=1$, $\theta=\pi/4$, $\delta=0.6$, $V=2$, and $\gamma=1$. }\label{Fig7}
\end{figure}

We consider the dynamical evolution of the master equation with the initial state taken as the ground state with the lowest real part of the spectra of the effective Hamiltonian $H_{\mathrm{eff}}^{(1)}$ with $L=8$, $N=4$, $\gamma=1$, and different $U_r$ under the PBC. For a trajectory that does not involve any loss event, along the trajectory, the initial state being the eigenstate of $H_{\mathrm{eff}}^{(1)}$ remains unchanged, and the corresponding physical quantities stay constant. When we include the effect of quantum jumps, the created holes scramble over time due to the two-particle loss. As concrete examples, we choose $U_r=5$ and $100$ to do our calculations. In the numerical simulation, we use $\mathcal{N}=10^4$ trajectories for $U_r=5$ and $\mathcal{N}=10^5$ trajectories for $U_r=100$. Figure \ref{Fig6} shows the dynamics of $\overline{\langle H_{\mathrm{eff}}^{(1)}(t)\rangle}$ for the dissipative eight-site systems in the present of quantum-jump events with $\delta=0.6$, $V=2$, $\theta=\pi/4$, and $\gamma=1$ under the PBC. Here, $\overline{\cdots}$ denotes a statistical average over different quantum trajectories and $\langle H_{\mathrm{eff}}^{(1)}(t)\rangle$ denotes the quantum-mechanical expectation values $H_{\mathrm{eff}}^{(1)}$ at time $t$. As shown in Figs. \ref{Fig6}(a1) and \ref{Fig6}(a2) for $U_r=5$, both the real and imaginary parts of $\overline{\langle H_{\mathrm{eff}}^{(1)}(t)\rangle}$ rapidly increase with time, compared to the case in the absence of quantum-jump events $\langle H_{\mathrm{eff}}^{(1)}(t)\rangle=-7.07 - 0.078 i$, denoted by the dashed lines. The result means that the system rapidly involves many more quantum-jump events with the increase of time. In the strong-repulsion limit $U_r=100$ shown in Figs. \ref{Fig6}(b1) and \ref{Fig6}(b2), we can see the values of $\langle H_{\mathrm{eff}}^{(1)}(t)\rangle$ approach the results in quantum trajectories without quantum-jump events in the short-time limits. With the increase of time, the probability of quantum-jump events occurring grows much slower than the in the $U_r=5$ case.

Both the initial states for $U_r=5$ and $U_r=100$ are Mott insulator states which present a charge density wave (CDW) structure and can be identified by the structure factor $\overline{S_k(t)}$, where
\begin{equation}
S_k=\frac{1}{L}\sum_{i,j=1}^{L}e^{ik(i-j)}[\langle n_i n_j\rangle-\langle n_i\rangle \langle n_j \rangle],
\end{equation}
with $k=2\bar{m}\pi/L,\bar{m}=0,1,\cdots,L$. The CDW phase can be characterized by the onset of the $k_F=2\pi N/L$ peak in the structure factor. In our case, the initial states are in the subspace of $N=4$, and the peak of $\overline{S_k(0)}$ is at $k=\pi$ for both cases, as shown by the dashed lines in Figs. \ref{Fig7}(a) and \ref{Fig7}(b). In the short-time limits ($t=5$, as shown in Fig. \ref{Fig7}), one can see that the system still presents the structure of a CDW.  For the $U_r=5$ case, the $k_F$ peak of $\overline{S_k(t)}$ decreases with the increase of time and finally vanishes, which suggests the instability of the Mott phase.
However, for $U_r=100$, even at $t=40$, the system is still in the CDW-dominated phase characterized by the $k_F$ peak, and the tiny decay of the values of $\overline{ S_{k_F}(t)}$ [see the inset of Fig. \ref{Fig7}(b)] implies that the probability of the quantum-jump events occurring is greatly suppressed and the Mott phase can still be observed.

According to the numerical calculations of the full master equation of the modulated Bose-Hubbard model with two-particle loss, we find that the quantum-jump term can be negligible in a sufficiently short time, which relies on the interaction strength.
While the system is unstable and quickly decays with time for the intermediate $U_r$ cases, the non-Hermitian TMI states are stable in the strong-repulsion limit.
The fact that the correlated topological states are maintained even in the presence of the quantum-jump term, suggests that the topological properties can be encoded in the effective non-Hermitian Hamiltonian in some parameter regions. We believe that our study will motivate further studies on the exploration of correlated topological properties in quantum open systems.

{\it Note added.} Recently, we became aware of the similar works of \cite{Danwei,TaoLiu}, which discuss the non-Hermitian topological Mott insulators in the interacting non-Hermitian Aubry-Andr\'{e}-Harper model with different focus issues.

\begin{acknowledgements}
	Z.Xu. is supported by the NSFC (Grant No. 11604188) and STIP of Higher Education Institutions in Shanxi under Grant No. 2019L0097. S.C. was supported by the NSFC (Grant No. 11974413) and the NKRDP of China (Grants No. 2016YFA0300600 and No. 2016YFA0302104). This work is also supported by NSF for Shanxi Province Grant No.1331KSC.
	
\end{acknowledgements}

\section*{Appendix A: Proof of many-body Chern number equalities}

In this appendix, we present the proof that the four many-body Chern numbers for the non-Hermitian cases give the same value. We generalize the proof in Ref. [\cite{ShenH}] for a single-particle case to many-body systems. We notice the fact that the Chern number is seen as an obstruction to a global gauge of the wave function in the parameter space $X=(\varphi,\theta)$ \cite{Kohmoto}, where $\varphi$ is a generalized boundary phase obtained by twisting the boundary condition and $\theta$ is a system parameter shown in the Hamiltonian (\ref{eq1}) in the main text. We consider a patch $P$ with a boundary $\partial P$ having circumference $C_L$ in the parameter space $X=(\varphi,\theta)$. In the $P$ regime, we choose a local gauge I and choose another local gauge II outside the $P$ regime, and corresponding eigenstates are denoted as $|\psi_{I}^{\alpha}\rangle$ and $|\psi_{II}^{\alpha}\rangle$, respectively, with $\alpha=\{R,L\}$. For the inner product $\langle \psi^{\alpha}(X)|\psi^{\alpha}(X)\rangle=1$, the two gauges on the boundary $\partial P$ are related by the gauge transformation $|\psi_{II}^{\alpha}(X)\rangle=e^{if(X)}|\psi_{I}^{\alpha}(X)\rangle$. However, for the inner product $\langle \psi^{\alpha}(X)|\psi^{\beta}(X)\rangle=1 \quad (\alpha\ne\beta)$, on the boundary the gauge transformations are written as $|\psi_{II}^{\alpha}(X)\rangle=r(X)e^{if(X)}|\psi_{I}^{\alpha}(X)\rangle$ and $|\psi_{II}^{\beta}(X)\rangle=e^{if(X)}/r(X)|\psi_{I}^{\beta}(X)\rangle$. Here, both $r(X)$ and $f(X)$ are continuous real functions. First, we consider $C^{RR}=C^{LR}$. For the Berry connection $A^{\alpha\beta}(X)=\langle\psi^{\alpha}(X)|\nabla_X\psi^{\beta}(X)\rangle$, the transformation law of the Berry connection can be written as
\begin{align}\label{eq8}
A_{II}^{RR}(X) &= A_{I}^{RR}+i\nabla_X f(X), \\
A_{II}^{LR}(X) &= A_{I}^{LR}+i\nabla_X f(X)+\frac{\nabla_X r(X)}{r(X)}.
\end{align}
The Berry curvature is the imaginary part of the curl of the Berry connection, and we can apply the Stokes theorem for the definition of the many-body Chern number,
\begin{align}\label{eq9}
C^{\alpha R} &=\frac{1}{2\pi}\int_{X} d\varphi d\theta B^{\alpha R}(\varphi,\theta) \notag  \\ &=\mathrm{Im}\left\{\oint_{\partial P}[A_{I}^{\alpha R}(X)-A^{\alpha_R}_{II}(X)]\cdot d\mathbf{l}\right\} \notag \\
&=\oint_{\partial P} \nabla_X f(X) \cdot d\mathbf{l}=f(C_L)-f(0),
\end{align}
where the result is independent of the choice of $\alpha$ and $r(X)$. For the $\alpha=L$ case, we have
\begin{equation}
\oint_{\partial P} \frac{\nabla_X r(X)}{r(X)}=\ln \frac{r(C_L)}{r(0)}=0.
\end{equation}
Thus, we can prove $C^{RR}=C^{LR}$. For $C^{LL}=C^{RL}$, we can apply the same process. And for $C^{RL}=C^{LR}$, we have $B^{RL}(\varphi,\theta)=B^{LR}(\varphi,\theta)$, so the integrals of those two Berry curvatures have the same values. Combining all the results, we finish the proof $C^{LL}=C^{RR}=C^{LR}=C^{LR}$.

\section*{Appendix B: Dynamical evolution of the master equation}

In this appendix, we describe the details of the method for solving the dynamical evolution of the full master equation described by Eq. (\ref{eq3}) with the Bose-Hubbard model $H_{\mathrm{BH}}$ subjected to a two-particle loss. To study the dynamics of a dissipative Hubbard model, we apply a quantum-trajectory method \cite{Dalibard,Daley,Plenio,Nakagawa2020,Ashida2020}, which involves rewriting the master equation as a stochastic average over individual trajectories. This method can avoid propagating a full density matrix in time and replaces the complexity with stochastic sampling. This change means if the Hilbert space has dimension $N_H$, then the density matrix propagates with the size $N_H^2$, whereas stochastic sampling of states requires the propagation of states of size $N_H$ instead. The penalty is the need to collect many samples for small statistical errors.

We follow a revised quantum-trajectory method that was originally proposed by Dum {\it et al.} \cite{Dum}, and the scheme takes the following form:

(1) Sample a random number $r_1$, uniformly distributed in the interval $0\le r_1 \le 1$.

(2) The wave function $|\tilde{\psi}(t)\rangle$ evolves under the nonunitary Schr\"{o}dinger equation $i\partial_{t}|\tilde{\psi}(t)\rangle=H_{\mathrm{eff}}^{(1)}|\tilde{\psi}(t)\rangle$. Numerically solve the equation $||\exp(-iH_{\mathrm{eff}}^{(1)}t_1)|\tilde{\psi}(\tau_0)\rangle||^2=r_1$ in order to find the time $t_1$ at which a loss event occurs. Here, $|\tilde{\psi}(0)\rangle$ is the initial state, and $H_{\mathrm{eff}}^{(1)}$ is the effective non-Hermitian Hamiltonian with two-particle loss under the PBC in the main text.

(3) The state is then computed numerically in the time interval $t \in[0,t_1]$ as
\begin{equation}
|\tilde{\psi}(t)\rangle=\frac{\exp(-iH_{\mathrm{eff}}t)|\tilde{\psi}(0)\rangle}{||\exp(-iH_{\mathrm{eff}}t)|\tilde{\psi}(0)\rangle||}.
\end{equation}
At time $t_1$, a quantum jump takes place, and the state $|\tilde{\psi}(t_1^+)\rangle$ is acted on by the quantum-jump operator $L_j$ with a particular $j$ based on the probabilities $\delta p_m \propto \langle \tilde{\psi}(t_1)|L^{\dagger}_{j}L_j | \tilde{\psi}(t_1) \rangle$ and then normalized:
\begin{equation}
|\tilde{\psi}(t_1^{+})\rangle=\frac{L_j|\tilde{\psi}(t_1^{-})\rangle}{||L_j|\tilde{\psi}(t_1^{-})\rangle||}.
\end{equation}

(4) After $t_1$, another random number $r_2$ is chosen, and the above procedure is repeated.

For a sufficiently large number $\mathcal{N}$ of samples of quantum trajectories, the density matrix of the solution of the full master equation is
\begin{equation}\label{eq12}
\rho(t)\approx \frac{1}{\mathcal{N}}\sum_{a=1}^{\mathcal{N}}|\tilde{\psi}_a(t)\rangle \langle \tilde{\psi}_a(t)|,
\end{equation}
where $|\tilde{\psi}_a(t)\rangle$ is the state along the $a$th quantum trajectory ($a=1,2,\cdots,\mathcal{N}$). When $\mathcal{N}\to \infty$, the result becomes an exact one.

\end{document}